\documentclass{sigchi-ext}
\usepackage[T1]{fontenc}
\usepackage{textcomp}
\usepackage[scaled=.92]{helvet} 
\usepackage{graphicx} 
\usepackage{balance}  
\usepackage{booktabs} 
\usepackage{ccicons}  
\usepackage{ragged2e} 

\def\plaintitle{Grounding Explainability} 
\def\emptyauthor{}
\def\plainkeywords{XAI; Global South; HCID; HCXAI; Human AI Interaction.}

\title{Grounding Explainability Within the Context of Global South in XAI}

\numberofauthors{4}
\author{%
  \alignauthor{%
    \textbf{Deepa Singh}\\
    \affaddr{Department of Philosophy } \\
    \affaddr{University of Delhi, India} \\
    \email{deepa.singh@gmail.com} } \alignauthor{%
    \textbf{Michal Slupczynski}\\
    \affaddr{RWTH Aachen University, Germany}\\
    \email{slupczynski@dbis.rwth-aachen.de} }\vfil \alignauthor{%
    \textbf{Ajit G. Pillai}\\
    \affaddr{Affetive Interactions Lab}\\
    \affaddr{The University of Sydney, Australia}\\
    \email{ajit.pillai@sydney.edu.au} }\alignauthor{%
    \textbf{Vinoth Pandian Sermuga Pandian}\\
    \affaddr{RWTH Aachen University}\\
    \affaddr{Aachen, Germany}\\
    \email{vinoth.pandian@rwth-aachen.de} }  }

\definecolor{linkColor}{RGB}{6,125,233}
\hypersetup{%
  pdftitle={\plaintitle},
  pdfauthor={\emptyauthor},
  pdfkeywords={\plainkeywords},
  bookmarksnumbered,
  pdfstartview={FitH},
  colorlinks,
  citecolor=black,
  filecolor=black,
  linkcolor=black,
  urlcolor=linkColor,
  breaklinks=true,
}

\CopyrightYear{2020}
\setcopyright{rightsretained}
\conferenceinfo{CHI'22, Workshop on Human-Centered Explainable AI (HCXAI): Beyond Opening the Black-Box of AI}{May  12--13, 2022, New Orleans, LA, USA}
\isbn{}
\doi{https://hcxai.jimdosite.com/}
\copyrightinfo{\acmcopyright}

\begin{document}

\maketitle

\justifying{} 

\begin{abstract}
  In this position paper, we propose building a broader and deeper understanding around Explainability in AI by ‘grounding’ it in social contexts, the socio-technical systems operate in. We situate our understanding of grounded explainability in the ‘Global South’ in general and India in particular and express the need for more research within the global south context when it comes to explainability and AI.
\end{abstract}



\section{Introduction}
Explainable AI (XAI) is a form of artificial intelligence (AI) that provides additional information to illustrate the decision process of a machine learning algorithm. An AI system can provide the end-users with explanations to increase interactivity by allowing them to re-enact and retrace AI/ML outcomes, for example, to verify results for plausibility \cite{holzinger2020measuring}. An extension to this definition is the concept of Responsible Artificial Intelligence (RAI), which describes an AI methodology with fairness, model explainability, and accountability at its core \cite{arrieta2020explainable}. Though there are clear definitions available in XAI, there are different perspectives on XAI depending on community context \cite{ehsan2021operationalizing} as a result of which there are various ways to operationalize and evaluate XAI technology. 

Explainability, interpretability, intelligibility, and transparency have been used interchangeably \cite{arrieta2020explainable,abdul2018trends,mohseni2021multidisciplinary,ehsan2021operationalizing}. 
To comprehend the formative and substantive human components of XAI systems, as well as the context of the many stakeholders who use the system, a contextual sociotechnical approach is required \cite{ehsan2020human}.
In this paper, we propose building a broader and deeper understanding of Explainability by ‘grounding’ it in the social contexts in which these socio-technical systems operate. 
We situate our understanding of grounded explainability in the ‘Global South’ in general and India in particular. 
The vast majority of the articles written on XAI were concentrated on the western counterparts \cite{islam2022systematic}. 
One of the major existing issues with XAI systems within contexts of the global south, especially in rural areas, is that most of the current systems are built with small amounts of explainability or with non-explainability, clearly showcasing the power structures between the developers of the system and marginalized communities using these systems.
Ehsan and Reidal \cite{ehsan2020human}, paint a picture of consequential technological systems used in society in different use cases are nested in social relationships. 
There is a pattern of neglecting this socially situated aspect within many AI and XAI contexts, especially when engineers are disconnected from the users, resulting in a half baked, partial and failed image of the involved information \cite{ehsan2020human}. 
Rather than focusing on how people interact with technologies, XAI should focus on what explainability, autonomy, and control mean to individuals with diverse backgrounds. 
Research and analysis of what these concepts imply are necessary to acquire a better grasp of how explainability might be fostered more successfully within these groups. 
In this ongoing investigation, we focus conducting research with a  transformative worldview and mixed methods research with quantitative and qualitative studies for ‘Grounding of explainability’ within the context of global south by centering the local communities is the approach we take in this paper to further understand what does such a grounding demand of us. 
How can we build the concept of Explainability or XAI in such a way that it leads to a tangible empowerment of local communities; as citizens, political participants, users and stakeholders? 
These are the questions we look to work on in the future. 
This workshop acts as a spring board for us, to have a discourse around this topic with fellow researches, gain feedback on our direction, resulting in a more informed research design for us to take forward. 
Our aim for the research to be conducted would be to look at what role XAI plays in enabling of trust and democratic participation by enabling communities in seeking accountability.
\section{Background}

There is a massive body of research that builds the case that though AI has benefits, it also has a panoply of potential harms associated with it. 
Research has also highlighted, that AI systems disproportionately harm the vulnerable, poor and the marginalized sections of society \cite{o2016weapons,broussard2018artificial,eubanks2018automating}. 
In addition, the development, research, and design of AI are mostly ``West''-centric / ``Global North''-centric \cite{sambasivan2021re,mhlambi2020rationality}. 
In this paper, we not only highlight the need to go beyond the established, Euro-America dominated the understanding of AI Explainability but we need to ground it in the real-time socio-political contexts such that it not only steers growth, development, and trust but also promotes democratic values while centering the worst-off of the society.

AI is currently under deployment in various sectors across India.
According to NITI Aayog’s (public policy think-tank of the Government of India) Approach Document for India, \textit{``[...] it is expected that AI usage will become ingrained and integrated with society. In India, large-scale applications of AI are being trialed every day across sectors.''} \cite{responsible}. 
There is evidence of AI being increasingly used in policing \cite{marda2020data}, development of the highly aspirational `smart city' projects, and even in welfare administration \cite{garg_2022}. 

There is a much-publicized government flagship project which aims to deploy AI and a host of other emerging technologies in India’s agricultural sector \cite{joshi}. 
In such a scenario, where incredible aspiration driving large scale AI deployments throughout the country, we wish to contextualize the pressing need for practical models / understanding / definitions of Explainability by ‘grounding’ its meaning and relevance in the well-being of the worst-off in the Indian society who are predominantly situated in rural India. 
The idea behind XAI is to render it transparent, open to scrutiny, and foster public trust. In this manner, XAI thus is also an effort in the direction of promoting and nurturing democratic values in societies, we argue. 
The fundamentals of XAI should therefore be grounded in the socio-political contexts of deployment to be enabling and have a tangible ground-level impact. 
The question that arises is what does Explainability mean in the context of the under-served, marginalized, and unlettered populations situated in the Global South, specifically, rural India? 
India, at present, does not even have a basic data protection framework. 
The ‘pacing gap’; between innovation and regulation is vast. 
Pre-existing institutional mechanisms for enabling transparency and accountability in sociotechnical systems are either weak or wholly absent. 
In want of adequate institutional mechanisms which can enable explainability and recourse post-hoc, we as the research community have to actively look into developing notions of explainability which could be grasped and exercised by the rural populace in India / the Global South. 
This is the central question that our research aims to tackles by proposing how and to what extent can we achieve such `grounding'.

\section{Research Direction}

In this research, we focus on looking at XAI in a transformative worldview. 
To tackle social injustice at its various levels, a transformative worldview emphasizes that academic inquiry must be interwoven with politics and a political transformation agenda, which also implies that the inquirer will ensure that the participants are not more marginalized as a result of the investigation \cite{mertens2010philosophy}.

We aim to conduct a mixed-method study with marginalized communities currently using AI systems with a participatory approach to understanding their problems with explainable AI and co-designing what explainability means to them when it comes to interacting with consequential AI systems. We plan to scope the XAI based on the problem domain where the tool disturbs our participant's livelihood --- for example, AI for loan calculation by banks or AI for seed distribution to farmers. From this formative research, we then plan to modify the XAI system and create a framework to provide an impactful solution to the target participants. Further, we will evaluate this system with the same community to assess its impact on the problem domain.

\section{About the authors}

\textbf{Deepa Singh}
is a PhD researcher at the Department of Philosophy, University of Delhi, where she is an Indian Council of Philosophical Research (ICPR) Fellow. Her research focuses on contextualising and (re)interpreting AI Ethics principles within the Indian context, and is situated at the intersections of Philosophy, HCI and AI. 

\textbf{Michal Slupczynski}
is a researcher and PhD student at the Chair for Computer Science 5 at RWTH Aachen University. He conducts research in infrastructures for multimodal learning environments. 

\textbf{Ajit G. Pillai}
is a researcher and PhD student at Design Lab in The University of Sydney. He regularly conducts research on the ethics of emerging technologies in HCI in industries such as healthcare. He regularly conducts his research with older adults and children within the health and well-being context. His PhD looks into the ethics of technology design.

\textbf{Vinoth Pandian}
is a HCI researcher and full stack web developer at MeetAnyway GmbH. He has a PhD in human centered artificial intelligence from RWTH Aachen university. His research focus is on the intersection of HCI and AI.

\balance{} 

\bibliographystyle{SIGCHI-Reference-Format}
\bibliography{sample}

\end{document}